\documentclass[prl,twocolumn,amssymb,amsmath]{revtex4-1}
\usepackage{graphicx}
\usepackage{epsfig,color}
\usepackage{amsmath,amssymb}
\usepackage{mathrsfs}

\usepackage{changes}

\begin{document}

\title{Superoscillations underlying remote state preparation for relativistic fields}

\date{\today}

\author{Ran Ber$^1$, Oded Kenneth$^2$ and Benni Reznik$^1$}

\begin{abstract}
\begin{center}
$^1$School of Physics and Astronomy, Raymond and Beverly Sackler
Faculty of Exact Sciences, Tel Aviv University, Tel-Aviv 6997801, Israel.

$^2$Department of physics, Technion 3200003 Haifa, Israel.
\end{center}
We present a physical (gedanken) implementation of a generalized remote state preparation of relativistic quantum field states for an arbitrary set of observers.
The prepared states are created in regions that are outside the future light-cone of the generating region.
The mechanism, which is based on utilizing the vacuum state of a relativistic quantum field as a resource, sheds light on the well known Reeh-Schlieder theorem, indicating its strong connection with the mathematical phenomenon of superoscillations.
\end{abstract}

\maketitle

\section*{I. Introduction}
Relativistic quantum field theory (QFT) provides a theoretical framework for unifying the classical theory of special relativity with the principles of quantum mechanics (QM).
From the standpoint of quantum information theory \cite{PreskillLN}, relativistic QFT has several appealing properties; quantum mechanical fields inherit the same causal structure of classical special relativity, and provide a concise formulation of the concepts of 'local observables' and 'local operations', which are fundamental for the study of entanglement in quantum information.

It is natural to expect that a quantum-relativistic framework would have significant implications to our understanding of quantum information \cite{Peres2004}, and vice versa, that the methods developed in quantum information could help improve our understanding of QFT.
Over the last decades there has been much research in this direction.
Following the pioneering work of Bohr and Rosenfeld \cite{Wheeler1983}, the measurability problems \cite{Aharonov1986,Sorkin1993,Beckman2001,Beckman2002,Groisman2002,Vaidman2003,Clark2010}, as well as relativistic quantum information tasks have been studied \cite{Pawlowski2009,Kent2011a,Kent2011,Kent2012,Kent2013,Hayden2013}.
An interesting observation in this context, is that relativistic QFT gives rise to entanglement between separated regions in space when the field is in the vacuum (zero-particle) state \cite{Reznik2003,Reznik2005,Verch2005,Silman2005,Massar2006,Silman2007,Botero2007,Orus2008,Hotta2008,Steeg2009,Marcovitch2009,Calabrese2009}.

\begin{figure}
\includegraphics[scale=0.34]{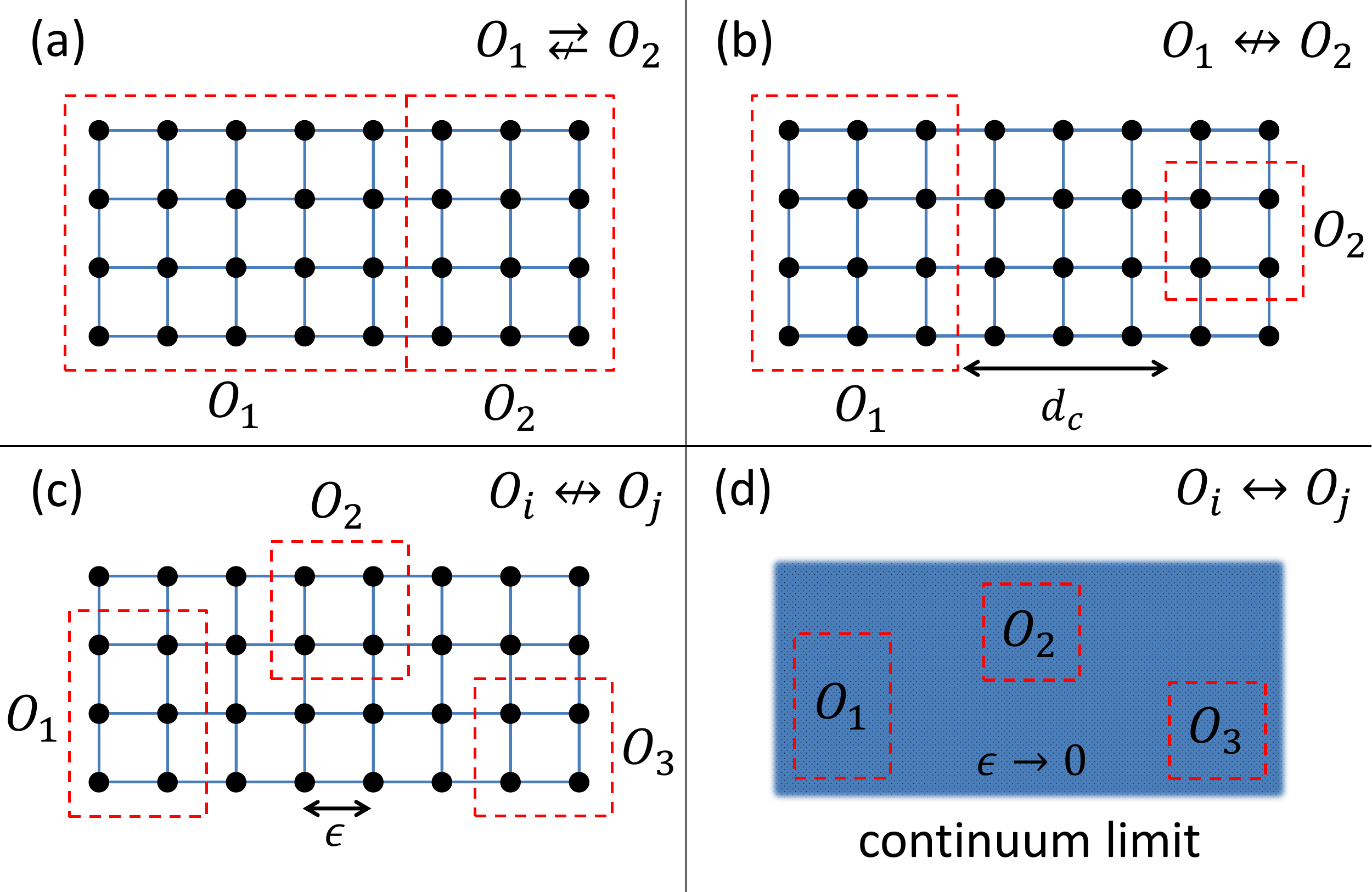}
\caption{
(Color online) Remote state preparation. (a) $O_1$ and $O_2$ are complementary regions of a system in a pure Gaussian state. It is possible to remotely prepare any state in $O_2$ by performing a measurement in $O_1$. (b) $O_1$ and $O_2$ are not complementary. In this case RSP is impossible. For $d>d_c$ the entanglement between $O_1$ and $O_2$ vanishes. (c) For several non-complementary regions $\left\lbrace O_k \right\rbrace$, RSP is impossible. (d) In the continuum limit, $\varepsilon \rightarrow 0$, the resulting QFT vacuum state can be used for RSP between any set of open regions $\left\lbrace O_k\right\rbrace$.}
\label{remote state generation QI}
\end{figure}

Since vacuum entanglement possesses this special feature, it is natural to ask whether it can be regarded as a resource for realizing new quantum information tasks.
It turns out that in the context of remote state preparation (RSP) \cite{Pati2000,Lo2000,Bennett2001} the answer is positive.
RSP is a process in which an observer prepares a desired quantum state in a remote system by performing a measurement on his own system. This process is possible due to shared entanglement between the systems.
A particular observer is said to have remotely prepared a certain desired state, if he is able to ascertain that for a particular measurement choice, and a particular outcome of this measurement, the remote system is in the required state.
The success probability in a single run can be small in general, however, it is required that for events with a successful measurement result, the remote state approaches the desired state with a fidelity arbitrarily close to one.
It is well known that RSP is possible when the Schmidt number of the initially shared (entangled) state is maximal.

Consider a lattice many body system, with a relativistic continuum limit, whose ground state is Gaussian (Fig. (1)). For two complementary regions $O_1$ and $O_2$, RSP from $O_1$ to $O_2$ can be realized provided that $\dim\left(\mathscr{H}_{O_1} \right)\geq\dim\left(\mathscr{H}_{O_2} \right)$ \footnote{A many body pure Gaussian state can be mapped by local operations at $O_1$ and $O_2$ to a pairwise entanglement form: A. Botero and B. Reznik, Phys. Rev. A 67, 052311 (2003) and Phys. Lett. A , 331, 39 - 44 (2004)} \cite{Botero2004} (Fig. (1a)).
For two non-complementary regions, RSP is generally not possible \footnote{This is because the two regions, along with the rest of the lattice, are in fact three complementary regions.}. In fact, for large enough separation, $d>d_c$, the two regions $O_1$ and $O_2$ become disentangled.
This is known as the phenomenon of ``sudden death of entanglement'' \cite{Yu2009} (Fig. (1b)).
For three complementary or non-complementary regions, RSP is impossible since that would require $\dim \left( \mathscr{H}_{O_i}\right)\geq \dim \left( \mathscr{H}_{O_j}\otimes \mathscr{H}_{O_l}\right)$ for any $i\neq j\neq l$, and this set of equations does not have a solution for finite dimensional Hilbert spaces $\mathscr{H}_{O_k}$ ($k\in \left\lbrace i,j,l \right\rbrace$) \cite{Clifton1998} (Fig. (1c)).
It is remarkable that in the limit $\varepsilon\rightarrow 0$ (where $\varepsilon$ denotes the lattice spacing), as the lattice approaches the continuum limit, RSP becomes possible in all the scenarios of Fig. (1a,b,c), as illustrated in Fig. (1d).

This follows from a fundamental, yet enigmatic, theorem about relativistic quantum field theories,
established long ago by Reeh and Schlieder \cite{Reeh1961,Schlieder1965,Haag1996}.
The theorem states that for any fixed open region $O_1$, acting on the vacuum (or on any other bounded energy state) by polynomials in the local operators corresponding to this region $\{ \phi(x)\;|\;x\in O_1\}$ generate a set of
states which is dense in the whole Hilbert space $\mathcal{H}$.
From an operational point of view this implies that by using local operations inside $O_1$ one may generate any
desired field state at some remote region(s) $\lbrace O_k \rbrace$ up to arbitrarily small infidelity.
As the required operations are typically not unitary, the process will involve post selection having certain
(non-zero) success probability.
The regions $\lbrace O_k \rbrace$ may remain throughout the process outside of the light-cone of $O_{1}$, hence its
outcome must be due to the pre-existing vacuum-correlations.

In this article we provide an operational method for applying RSP in relativistic QFT. This method can be regarded as a constructive proof of the Reeh-Schlieder theorem, which has been deduced in the abstract framework of algebraic QFT.

\begin{figure}
\includegraphics[scale=0.34]{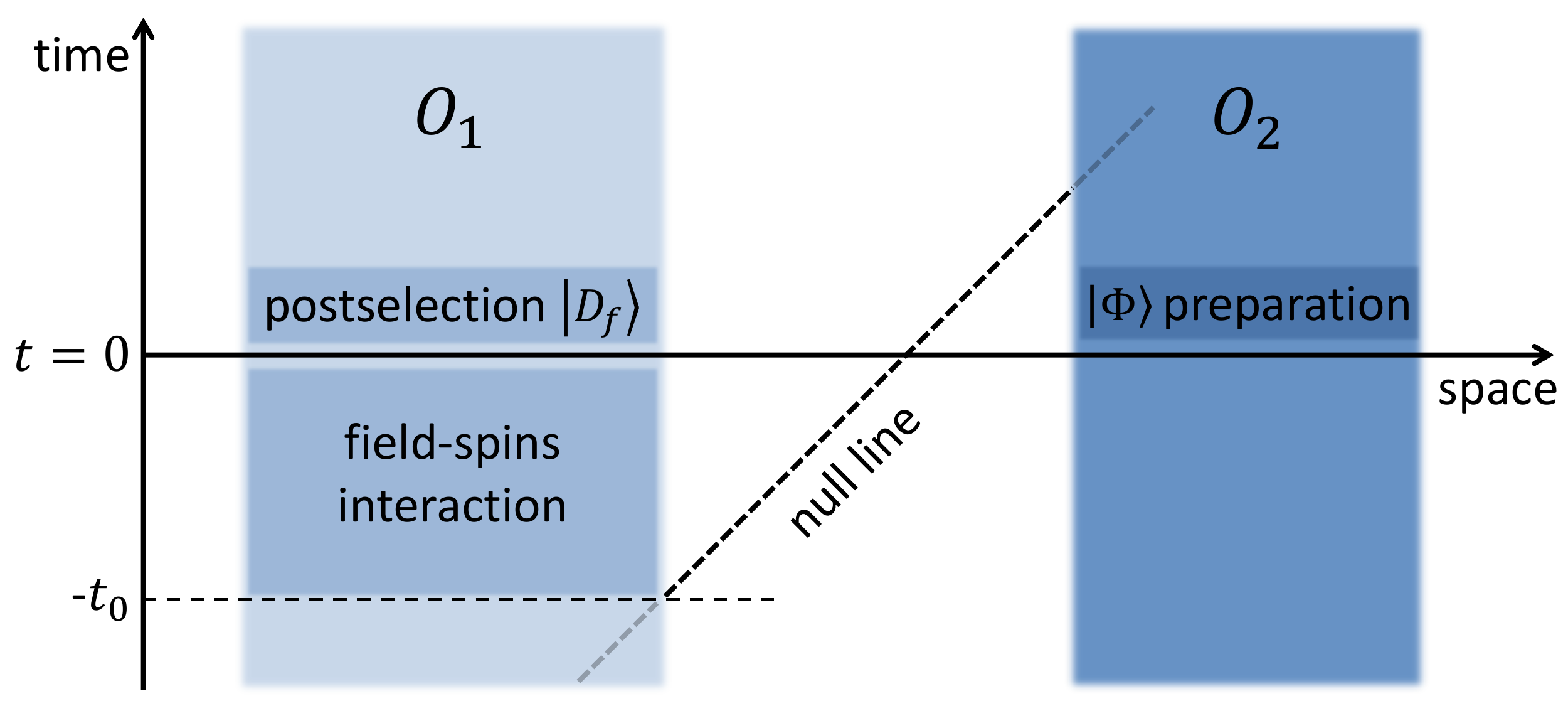}
\caption{(Color online) By interacting the field and the spins at $O_1$ in the time interval $-t_0<t<0$, one can prepare a state $\left|\Phi\right\rangle$ which is arbitrarily
close to a desired state $\left|\Psi\right\rangle$, located in $O_2$, even if $O_1$ and $O_2$ are causally disconnected throughout the process.}
\label{remote state generation}
\end{figure}

We employ the following scheme, as depicted schematically in Fig. (\ref{remote state generation}).
Consider two (or more) regions in space. In the ``generating'' region, $O_{1}$, a set of localized ``spin'' detectors \cite{Unruh1976} are arranged at specific positions $\bold{r}_i\in O_1$ ($i=1,...,N$).
The interaction between the spins and a relativistic field is turned on during $-t_{0}<t<0$; otherwise they remain decoupled from the field.
Relativistic causality then guarantees that by setting $t_0$ to be sufficiently small, certain ``remote'' regions $\lbrace O_k \rbrace$ ($k\ge2$) will remain causally disconnected from the spins and the field in $O_1$ up to $t=0$.
At $t=0$, once the spins are again decoupled from the field, we can postselect them to a state $|D_f\rangle$ \footnote{Since the postselection entangles the spins in a general way, this requires time that scales like the typical size of the region}.
As in the ordinary RSP scheme, while the unconditional local state at the remote region has not been changed (and so causality has not been conflicted), the conditional state has been modified to a pure state $\left\vert\Phi\right\rangle$.
This state can be guaranteed to be arbitrarily close to any desired pure state $|\Psi\rangle$.

More formally, we can describe the process as follows:
Given a field state, $|\Psi \rangle$, $t_0$ and $\eta>0$, we find a set of $N$ spins at $\bold{r}_i\in O_1$, certain local spins-field interactions for $-t_0<t<0$ and a spins' state $|D_f\rangle$, which at $t=0$ can be postselected with probability $p(\eta)>0$.
This particular postselection generates a field state $|\Phi\rangle$, which satisfies $|\langle \Psi|\Phi \rangle| >1-\eta$.

This paper is organized as follows. In section II we describe a general method for preparing field states by coupling the field to spins. In section III we present the superoscillations that are used in order to remotely prepare field states. In section IV we generalize the process for the generation of arbitrary field states in $d+1$ dimensions, and in section V we discuss the success probability and fidelity of the process. The paper also contains an appendix, expanding on the generation of arbitrary field states in $1+1$ dimensions.

\section*{II. Scheme for field state preparation}
We begin by considering a single spin at $\bold {r}=\bold {r}_1$ interacting with a Klein-Gordon field.
The spin-field interaction is taken to be
\begin{equation}\label{hint}
H_{\text{int}}=\lambda\left(\sigma_{+}\epsilon\left(t\right)+
\sigma_{-}\epsilon^{*}\left(t\right)\right)\phi\left(\bold{r}_{1}\right),
\end{equation}
where the complex window function, $\epsilon(t)$, is non-vanishing only for $-t_{0}<t<0$ and $\lambda$ is a small coupling constant.
Here the spins are modeled as non-relativistic first quantized objects.
Within a fully second quantized framework, one needs to describe them in terms of fields \cite{Unruh1976}.
However when the spins' mass is taken to be much larger than the typical frequencies of $\epsilon(t)$, pair creation and recoil effects are negligible \cite{Parentani1995,Reznik1998} and the interaction term reduces to Eq. (\ref{hint}).
Therefore any $\epsilon(t)$ is allowed given a sufficiently large spins' mass.

The spin-field initial state is $\left|d,\Phi\right\rangle_{t=-t_0}=\left|\downarrow,0\right\rangle$, where $\left|\downarrow\right\rangle $ is the ground state of the spin
and $\left|0\right\rangle $ denotes the field's vacuum state.

The interaction with the field leads, to first order in $\lambda$, to the state
\begin{equation}
\left|d,\Phi\right\rangle_{t=0}
 =\left|\downarrow,0\right\rangle
-i\lambda\int_{-t_{0}}^{0}dt\epsilon\left(t\right)e^{i\Omega t}\phi
\left(\bold{r}_{1},t\right)\left|\uparrow,0\right\rangle, \label{psi final}
\end{equation}
where $\phi$ is the field operator in the interaction picture and $\Omega$ is the energy gap of the spin's free Hamiltonian.

By measuring the spin we project the field, {\it conditionally}, to a particular state.
If the spin is found in the $\sigma_z=-1$ state, the field's state returns to the vacuum.
However, if the outcome is $\sigma_z=1$, the field's state will be modified into $\left\vert\Phi\right\rangle\propto \int_{-t_{0}}^{0}dt\epsilon\left(t\right)e^{i\Omega t}\phi\left(\bold{r}_{1},t\right)\left\vert 0\right\rangle$.
To illustrate the effect of our procedure on the field, let us then consider for simplicity the 1+1 dimensional case.
By projecting Eq. (\ref{psi final}) on the state $\phi\left(x_{1}+L',0\right)\left\vert \uparrow,0\right\rangle$ (where $L'$ is the position relative to $x_1$), we obtain the amplitude ${\cal A}_\uparrow \left(L';\Omega, \lbrace \epsilon\left(t\right)\rbrace\right)\!=\!\langle \uparrow, 0| \phi(x_1\!+\!L',0)|d,\Phi \rangle_{t=0}$, which can be expressed as
\begin{equation}\label{Aup}
{\cal A}_\uparrow \propto\int_{-\infty}^{\infty}dk\int_{-t_{0}}^{0}dt\epsilon\left(t\right)\frac{1}
{\omega\left(k\right)}e^{i\left(\omega\left(k\right)+\Omega\right)t}e^{ikL'},
\end{equation}
where $m$ is the mass of the field and $\omega\left(k\right)=\sqrt{m^2+k^2}$ ($\hbar=c=1$). Consider the condition
\begin{equation}\label{Aup=D}
{\cal A}_\uparrow(L';\Omega, \lbrace \epsilon\left(t\right)\rbrace) \propto D\left(L'-L,0\right)+D\left(L'+L,0\right),
\end{equation}
where $D\left(x-x',t-t'\right)=\left\langle 0|\phi(x',t')\phi(x,t)|0\right\rangle$ is the free Klein-Gordon propagator (reflection symmetry around $x=x_1$ is due to the absence of directional preference in a single point-like coupling case).
If this condition is met, then after postselecting spin ``up'', the vacuum state has changed into $(\phi(x_1+L)+\phi(x_1-L)) |0\rangle$.
This implies a deterministic (conditional) operation of applying the field operator $\phi(x=x_1\pm L,t=0)$ to the vacuum state.
Defining $\tilde{\epsilon}\left(\omega+\Omega\right)\equiv\int_{-t_{0}}^{0}dt\epsilon\left(t\right)e^{i\left(\omega+\Omega\right)t}$, and comparing equations (\ref{Aup}) and (\ref{Aup=D}), lead to the condition
$\tilde{\epsilon}_\text{des}\left(\omega+\Omega\right)\propto
\cos\left(\sqrt{\omega^{2}-m^{2}}L\right)$, where $\tilde{\epsilon}_\text{des}\left(\omega+\Omega\right)$ is the desired form of $\tilde{\epsilon}\left(\omega+\Omega\right)$.
For $L>t_0$, i.e., for points outside of the future light-cone of the spin, we observe that $\tilde\epsilon_\text{des}(\omega+\Omega)$ has significant Fourier components which oscillate, in frequency space, at ``frequencies'' $t<-t_0$ and $t>0$, while $\epsilon\left(t\right)$ only has support in $\left[-t_{0},0\right]$.
The standard basic frequency-time relations of Fourier transforms suggest that the above relation cannot be satisfied.

\section*{III. The necessity of superoscillations}
In order to circumvent the above problem, we have to make use of special tailored functions that oscillate faster than their fastest Fourier component. This type of oscillations is called ``superoscillations'' \cite{Aharonov1988}.
Superoscillatory functions have been extensively studied recently, both theoretically \cite{Berry1994a,Berry1994,Reznik1997,Rosu1997,Kempf2000,Ferreira2002,Aharonov2002,Kempf2004,Berry2006,Ferreira2006,Berry2009}
and experimentally \cite{Rogers2012,Wong2013}.

Superoscillations come with a price:
since superoscillations are due to destructive interference, they are always accompanied by exponentially larger amplitudes somewhere outside the superoscillatory region.
Fortunately, in relativistic QFT models, the energy is bounded from below.
In our case $\omega'\equiv\omega(k)+\Omega-m>\Omega>0$, hence one can select a proper superoscillatory function $\tilde{\epsilon}\left(\omega'\right)$ which manifests its exponential growth strictly outside the physical range of the frequency.
The amplitude of the function in the superoscillatory domain will, however, remain small. This, in turn, will cause the exponential decay of the success probability with the distance.
Another difficulty regarding superoscillations is that these functions can superoscillate in an arbitrarily large, but not infinite domain.
Therefore, there must also be a physical non-superoscillatory domain at $\omega'>\omega_c$ for some $\omega_c$.
This domain (unlike the non-physical one) will not be exponentially
amplified. Its effect will therefore only be to add amplitude for regular particle creation inside the causal light-cone.
This contribution can be compensated by destructively interfering it with ordinary processes amplitude as long as $\tilde{\epsilon}$ decays fast enough as $\omega'\rightarrow \infty$.
Furthermore, since the superoscillatory domain is bounded, the condition described in Eq. (\ref{Aup=D}) cannot be satisfied exactly. However, one can get arbitrarily close to satisfying this condition by increasing the superoscillatory domain.

\begin{figure}
\includegraphics[scale=0.34]{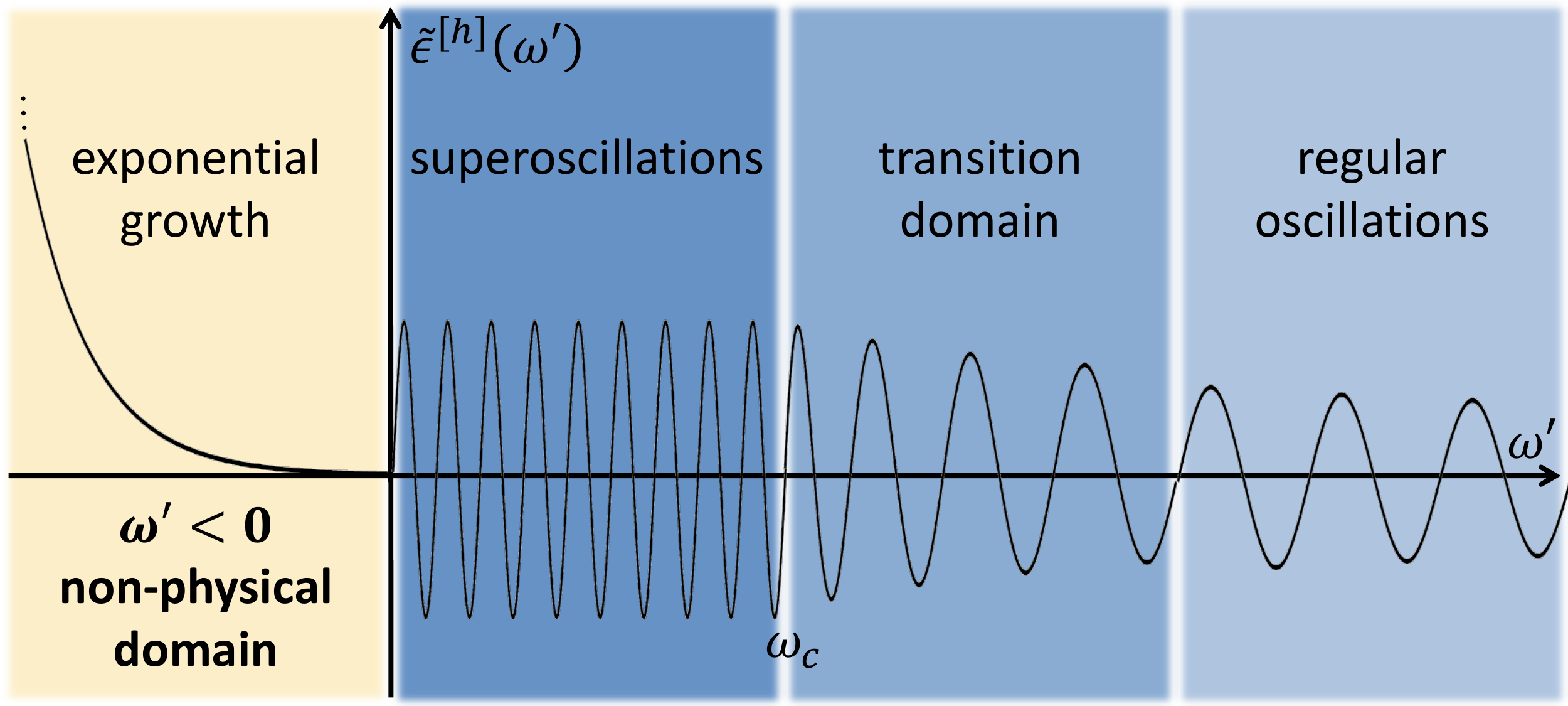}
\caption{(Color online) A schematic plot of the superoscillatory function that we use: the function obtains its exponential growth at the non-physical domain $\omega'<0$. At $0<\omega'<\omega_c$ the function superoscillates, and around $\omega_c$ it gradually obtains regular (slower) oscillations. The function decays as $\omega' \rightarrow \infty$ in order for the contribution beyond the superoscillatory domain to be compensatable.}
\label{superoscillatory functions}
\end{figure}

Before we proceed, we note that if one manages to find superoscillatory functions $\tilde{\epsilon}^{[h]}\left(\omega'\right)$ which oscillate like $\exp\left(i\omega' t'\right)$ for an arbitrary $t'<-t_0$ or $t'>0$, he would be able to use them (combined with regular oscillating functions having $-t_0<t'<0$) in order to assemble the desired function by a Fourier transform
\footnote{The Fourier transform of the desired function does not typically have a compact temporal support, however, an approximated Fourier transform, truncated at an arbitrary large $T$, will suffice.}.
In Fig. (\ref{superoscillatory functions}) we present a sketch of the superoscillatory function that we seek.

\begin{figure}
\includegraphics[scale=0.34]{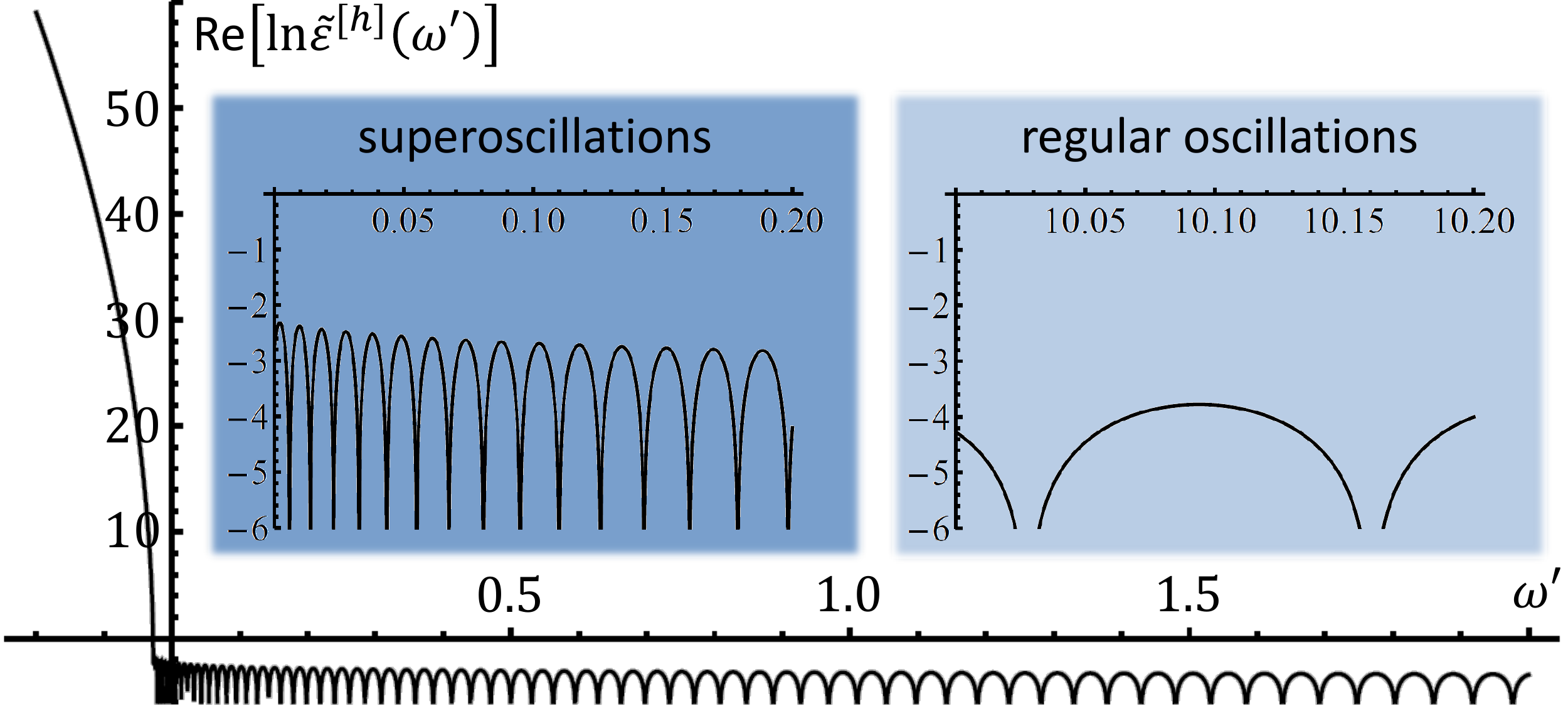}
\caption{(Color online) A logarithmic plot of the superoscillatory function presented in Eq. (\ref{epsilon(omega)}) with the parameters $\Delta=0.1$, $\delta=0.2$ $A=7.5$ and $t_0=1$.
Note the exponential growth at $ \omega'<0$.
The superoscillations at small $\omega'$ and the regular (slower) oscillations at larger $\omega'$ are shown in the insets.}
\label{superoscillatory function}
\end{figure}

We shall now proceed by finding such functions. Consider the following function \cite{Berry1994,Reznik1997}:
\begin{equation}
\tilde{\epsilon}^{[h]}\left(\omega'\right)=\frac{\Delta}{2\delta\sqrt{2\pi}}
\int_{0}^{2\pi}d\alpha e^{i\omega't_{0}\left(\frac{\cos\alpha-1}{2}\right)}e^{\frac{i}{\delta^{2}}
\cos\left(\alpha-iA\right)},\label{epsilon(omega)}
\end{equation}
where $\Delta$, $\delta$ and $A$ are some constants. A logarithmic plot of this function is presented in Fig. (\ref{superoscillatory function}).
While $\epsilon^{[h]}\left(t\right)$ has compact temporal support (since $t=t_0\left(\cos\alpha-1\right)/2\in \left[-t_0,0\right]$), we will now show that it can oscillate in $\omega$ space arbitrarily fast.
Performing the integration explicitly we obtain
\begin{eqnarray}
\tilde{\epsilon}^{[h]}\left(\omega'\right)&=&\frac{\Delta\sqrt{\pi}}{\sqrt{2}\delta} e^{-\frac{1}{2}i \omega' t_0}\nonumber \\
&&\!\times J_{0}\!\left(\frac{1}{\delta^{2}}\!\sqrt{1\!+\!\delta^{2}\omega't_{0}\cosh\left[A\right]\!+\!\frac{1}{4}\delta^{4}\omega'^{2}t_{0}^{2}}\right)\!.\label{eJ}
\end{eqnarray}
Using the asymptotic form of the Bessel function for $\delta\ll1$ we get
\begin{widetext}
\begin{equation}
\tilde{\epsilon}^{[h]}\left(\omega'\right)\cong \frac{\Delta}{\left(1+\delta^{2}\omega't_{0}\cosh\left[A\right]+\frac{1}{4}\delta^{4}\omega'^{2}t_{0}^{2}\right)^{\frac{1}{4}}} e^{-\frac{1}{2}i \omega' t_0}
\cos\left(\frac{1}{\delta^{2}}\sqrt{1+\delta^{2}\omega't_{0}\cosh\left[A\right]+\frac{1}{4}\delta^{4}\omega'^{2}t_{0}^{2}}-\frac{\pi}{4}\right).
\end{equation}
\end{widetext}
In order to obtain the superoscillatory domain, $\left[0,\omega_c\right]$, we take $\delta^{2}\ll\frac{1}{\omega_c t_{0}\cosh\left[A\right]}$.
Then, for $\omega'\in\left[0,\omega_c\right]$ this function reduces to
\begin{equation}
\tilde{\epsilon}^{[h]}\left(\omega'\right)\cong\Delta e^{-\frac{1}{2}i\omega' t_0}\cos\left(\frac{1}{\delta^{2}}+\frac{1}{2}\omega't_{0}\cosh\left[A\right]-\frac{\pi}{4}\right).
\end{equation}
One may fix the phase by choosing $\delta^{-2}=2\pi m+\pi/4$ where $m\gg1$ to get
\begin{equation}\label{dd}
\tilde{\epsilon}_1^{[h]}\left(\omega'\right)\cong\Delta e^{-\frac{1}{2}i\omega' t_0} \cos\left(\frac{1}{2}\omega't_{0}\cosh\left[A\right]\right)
\end{equation}
and $\delta^{-2}=2\pi m-\pi/4$ to get
\begin{equation}
\tilde{\epsilon}_2^{[h]}\left(\omega'\right)\cong\Delta e^{-\frac{1}{2}i\omega' t_0} \sin\left(\frac{1}{2}\omega't_{0}\cosh\left[A\right]\right).
\end{equation}
Therefore,
\begin{equation}
\tilde{\epsilon}_3^{[h]}\left(\omega'\right)\equiv\tilde{\epsilon}_1^{[h]}\left(\omega'\right)\pm i\tilde{\epsilon}_2^{[h]}\left(\omega'\right)\cong\Delta e^{\frac{1}{2}i\omega' t_0\left(\pm\cosh\left[A\right]-1 \right)}.
\end{equation}
This function oscillates in $\omega$ space at ``frequency'' $t'=\frac{1}{2}t_{0}\left(\pm\cosh\left[A\right]-1\right)$, therefore this segment is referred to as the superoscillatory domain.
By increasing $A$ we can set these oscillations to be arbitrarily fast.
The superoscillatory domain is finite, however, by decreasing $\delta$ it could be set to be arbitrarily large.

This function gets exponentially amplified at $\omega'< -2e^{-A}/(t_0\delta^2)<0$, where the argument of the Bessel function becomes imaginary. However, since $\Omega\geq 0$ the growth corresponds to $\omega<m$, which is a non-physical domain.
Beyond the superoscillatory domain, the function gradually obtains regular (slower) oscillations, and in the limit $\omega'\gg\omega_c$ it becomes $\tilde{\epsilon}_1^{[h]}\left(\omega'\right)\sim\frac{\Delta}{\sqrt{\omega'}}e^{-\frac{1}{2}i\omega't_0}\cos\left(\frac{1}{2}\omega't_{0}\right)$. The slow decay is related to the fact that $\epsilon^{[h]}(t)$ is not smooth at $t=-t_0,0$ (see Eq. (\ref{e(t)})).
In order to induce a faster decay we convolute $\epsilon^{[h]}(t)$ with a smooth function $h(t)$ having a very small temporal support.
This amounts to replacing $\tilde{\epsilon}^{[h]}\left(\omega'\right)$ by $\tilde{\epsilon}^{[h]}\left(\omega'\right)\tilde{h}\left(\omega'\right)$.
Assuming $h\left(t\right)$ is differentiable $n$ times ensures that the new $\tilde{\epsilon}^{[h]}\left(\omega'\right)$ decays like $\omega'^{-\left(n+\frac{1}{2}\right)}$ outside the superoscillatory domain.
For $n>\frac{1}{2}\left(d-2\right)$ it decays fast enough for $\tilde{\epsilon}^{[h]}\left(\omega'\right)$ to be normalizable. Once $\tilde{\epsilon}^{[h]}\left(\omega'\right)$ is normalizable, the contribution beyond the superoscillatory domain can be compensated by destructively interfering it with ordinary processes amplitude.

We can use a combination of such superoscillatory functions,
each with a different $t'$, in order to generate the window function
\begin{equation}
\tilde{\epsilon}\left(\omega'\right)  =  \int_{-T}^{T}dt'\tilde{\epsilon}_{3}^{[h]}\left(\omega';t'\right)\epsilon_\text{des}\left(t'\right),
\end{equation}
where $T=\frac{1}{2}t_{0}\left(\cosh\left[A_\text{max}\right]+1\right)$. In the limits $T\rightarrow\infty$ and $\delta\rightarrow 0$ we get $\tilde{\epsilon}\left(\omega'\right)\rightarrow \tilde{\epsilon}_\text{des}\left(\omega'\right)$
in the segment $\omega'\in\left[0,\omega_{c}\right]$. This is while the actual window function, $\epsilon\left(t\right)$, and the desired window function, $\epsilon_\text{des}\left(t\right)$, are very different: $\epsilon\left(t\right)$ has temporal support only in $\left[-t_0,0\right]$, while $\epsilon_\text{des}\left(t\right)$ might have an arbitrarily large temporal support.

\section*{IV. Generation of arbitrary states}
Let us now proceed by using the above results to demonstrate the generation of arbitrary field states in $d+1$ dimensions.
We shall start by generating a one--particle spherical symmetrical state around a single spin. Next, we will generate an arbitrary one--particle state using an array of spins, and finally we will generalize this process to many--particle states.

\subsection*{A. Spherical symmetrical one--particle states}
Using a single spin at $\bold{r}_1=0$, we have shown in section II that one obtains the state
$|\Phi\rangle\propto\int_{-t_{0}}^0 dt\epsilon\left(t\right)e^{i\Omega t}\phi(0,t)|0\rangle$.
This is clearly a spherically symmetric one--particle state.
It is therefore of the general form $\left|\Phi\right\rangle=\int d^d\bold{r} f\left(r\right)
\phi(\bold{r},0)|0\rangle$ for some radial weight function $f\left(r\right)$.
Substituting the standard expansion of $\phi$ in terms of creation and annihilation
operators we obtain
\begin{eqnarray}
|\Phi\rangle&\propto&\int_{-t_{0}}^0 dt\epsilon\left(t\right)e^{i\Omega t}\phi(0,t)\left|0\right\rangle\nonumber \\
&=&\int {d^d\bold{k}\over(2\pi)^d}\tilde{\epsilon}(\omega(k)+\Omega)){1\over\sqrt{2\omega(k)}}\left|\bold{k}\right\rangle.
\end{eqnarray}
We would like this to coincide with the state
\begin{eqnarray}
\left|\Psi\right\rangle&=&\!\int\! d^d\bold{r}\! f\left(r\right)\phi(\bold{r},0)\left|0\right\rangle\nonumber \\
&\!\propto\!&\!\int_{0}^\infty \!dr\! f\!\left(\!r\!\right)\!\int\! {d^d\bold{k}\over(2\pi)^d}
k\!\left({2\pi r\over k}\right)^{d/2}\!J_{\frac{d\!-\!2}{2}}(kr){1\over\sqrt{2\omega\!\left(\!k\!\right)}}\left|\bold{k}\right\rangle\!,\nonumber \\
&&
\end{eqnarray}
where $J$ stands for Bessel function. This leads to
\begin{equation}
\tilde{\epsilon}_\text{des}(\omega(k)+\Omega)\propto
\int_0^\infty dr f\left(r\right) k\left({2\pi r\over k}\right)^{d/2}J_{\frac{d\!-\!2}{2}}(kr).
\end{equation}

\subsection*{B. Arbitrary one--particle states}
In order to generate field states which are not spherically symmetric, we replace the single spin by an array of (possibly a large number of) such spins, all located inside the region $O_1$.
Expanding perturbatively up to the first order in $\lambda$, the most general field state generated by $N$ spins is
\begin{eqnarray}
\left|d,\Phi\right\rangle &=&\left|\downarrow\downarrow\ldots\downarrow,0\right\rangle + \nonumber\\
&&\lambda\!\underset{i}{\sum}\!\int_{-t_{0}}^0 dt\epsilon_i\left(t\right)e^{i\Omega t}\phi(\bold{r}_i,t)\left|\lbrace i \rbrace,0\right\rangle\!+\!\mathcal{O}\left(\lambda^{2}\right), \nonumber\\
&&
\end{eqnarray}
where $\left|\lbrace i \rbrace\right\rangle$ denotes a state in which the $i$'th spin points ``up'' and the remaining spins point ``down''.
By postselecting the spins to the state $\left|d_f\right\rangle =\sum\alpha_{i}^{*}\left|\lbrace i \rbrace\right\rangle$ we obtain
\begin{equation}
\left|\Phi\right\rangle =\underset{i}{\sum}\alpha_{i}\int_{-t_{0}}^0 dt\epsilon_i\left(t\right)e^{i\Omega t}\phi(\bold{r}_i,t)\left|0\right\rangle.
\end{equation}
For convenience, we shall imagine a continuous ``spin distribution''.
This can be approximated arbitrarily well by a discrete distribution consisting of a very large yet finite number of spins.
The resulting field state is then
\begin{equation}\label{tyg}
\left|\Phi\right\rangle =\underset{O_1}{\int} d^d\bold{r}\alpha\left(\bold{r}\right)\int_{-t_{0}}^0 dt\epsilon\left(t\right)e^{i\Omega t}\phi(\bold{r},t)\left|0\right\rangle,
\end{equation}
where we have assumed that all spins are coupled to the field using the same window function.

In the following we shall assume for simplicity $3+1$ dimensions.
Generalizing to $d+1$ dimensions is, however, straight forward.
As any state can be expanded in spherical harmonics it will be enough to consider states having their angular dependence given by some fixed $Y_{lm}(\hat{\bold{r}})$.
In order to achieve this we choose the following ``weight'' function $\alpha\left(\bold{r}\right)=a_0^{-2}Y_{lm}(\hat{\bold{r}})\delta(r-a_0)$.
We then have
\begin{equation}
|\Phi\rangle\!=\!\frac{1}{a_0^2}\!\int_{-t_0}^0\!dt\epsilon\left(t\right)\!e^{i\Omega t}\!\int \!d^3\bold{r} \delta(r\!-\!a_0)\!Y_{lm}\!(\hat{\bold{r}})\phi(\bold{r}\!,\!t)|0\rangle.
\end{equation}
A straight forward calculation shows that
\begin{eqnarray}
&&\!\int\! d^3\bold{r} F_l(r)Y_{lm}(\hat{\bold{r}})\phi(\bold{r},t)|0\rangle\nonumber \\
&=&\!\int\! {d^3\bold{k}\over\sqrt{2\omega_k}}\!\int_0^\infty \!F_l(r)r^2dr\!\int d^2\Omega_{\bold{r}} e^{i\omega_k t}e^{i\bold{k}\cdot\bold{r}}Y_{lm}(\hat{\bold{r}})|\bold{k}\rangle\nonumber \\
&=&\!\int\! {d^3\bold{k}\over\sqrt{2\omega_k}}\!\int_0^\infty \!F_l(r)r^2dr e^{i\omega_k t} 4\pi i^l j_l(kr)Y_{lm}(\hat{\bold{k}})|\bold{k}\rangle\!,
\end{eqnarray}
for an arbitrary radial function $F_l\left(r\right)$. Thus, in particular, we find
\begin{eqnarray}
\left\vert\Phi\right\rangle&=&\!\!\int_{-t_0}^0\!\!dt\epsilon\left(t\right)e^{i\Omega t} \!\!\int \!\!{d^3\bold{k}\over\sqrt{2\omega_k}}
 e^{i\omega_k t} 4\pi i^l j_l(ka_0)Y_{lm}(\hat{\bold{k}})|\bold{k}\rangle\nonumber \\
&=&4\pi \!\int \!{d^3\bold{k}\over\sqrt{2\omega_k}} \tilde{\epsilon}\left(\Omega+\omega_k\right) i^l j_l(ka_0)Y_{lm}(\hat{\bold{k}})|\bold{k}\rangle,
\end{eqnarray}
while the desired final state is by the same calculation
\begin{eqnarray}
\left\vert\Psi\right\rangle & = & \int d^3\bold{r} F_l(r)Y_{lm}(\hat{\bold{r}})\phi(\bold{r},0)|0\rangle \nonumber\\
& = & 4\pi\!\int\! {d^3\bold{k}\over\sqrt{2\omega_k}}\!\int_0^\infty\! F_l(r)r^2dr
 i^l j_l(kr)Y_{lm}(\hat{\bold{k}})|\bold{k}\rangle. \nonumber\\
 &&
\end{eqnarray}
Therefore, to obtain $\left|\Phi\right\rangle=\left|\Psi\right\rangle$ we need
\begin{equation}
\tilde{\epsilon}_\text{des}\left(\Omega+\omega_k\right) j_l\left(ka_0\right)=\int_0^\infty F_l\left(r\right)r^2dr j_l\left(kr\right).
\end{equation}
Taking for example $F_l(r)=\delta(r-R)$ (with $R>ct_0+a_0$) we find the condition
\begin{equation}
\tilde{\epsilon}_\text{des}\left(\Omega+\omega_k\right)\sim j_l\left(kR\right)/j_l\left(ka_0\right).
\end{equation}
To avoid possible singularities of the r.h.s one has to set $a_0$ such that $a_0 k\left(\omega_c\right)<Z_{l,1}$, where $Z_{l,1}$ is the first non-trivial zero of the $l$'th spherical Bessel Function.
The limit $a_0\rightarrow 0$ actually corresponds to a single effective interaction with a high multipole of the field $\phi$.

\subsection*{C. Arbitrary states}
So far we have demonstrated the generation of arbitrary one--particle field states of the form
\begin{equation}
\left\vert\Psi\right\rangle = \int d^d\bold{r} F\left(\bold{r}\right)\phi(\bold{r},0)|0\rangle.
\end{equation}
In order to generate an arbitrary $M$--particle product state
\begin{equation}
\left\vert\Psi\right\rangle = \prod_i\int d^d\bold{r}_i F_i\left(\bold{r}_i\right)\phi(\bold{r}_i,0)|0\rangle,
\end{equation}
one has to use $M$ such spin arrays and postselect them in the state
\begin{equation}
\left|D_f\right\rangle= \left|d_{f}^{\left[1\right]}d_{f}^{\left[2\right]}\ldots d_{f}^{\left[M\right]}\right\rangle.
\end{equation}
In order to avoid ordering issues, one may assume the spin arrays are mutually casually disconnected throughout the interaction duration. In order to generate an arbitrary (usually entangled) $M$--particle state, one would have to postselect the spins in the state
\begin{equation}
\left|D_f\right\rangle= \underset{ab...m}\sum C_{ab...m}\left|d_{f,a}^{\left[1\right]}d_{f,b}^{\left[2\right]}\ldots d_{f,m}^{\left[M\right]}\right\rangle.
\end{equation}
The generalization to a superposition of field states with different numbers of particles is straightforward.

\section*{V. Success probability and fidelity}
Vacuum entanglement between separated regions of space-time decays exponentially with the separation.
We therefore expect that the chance to successfully generate a field state $\phi\left(x_{1}+L\right)|0\rangle$ far away from a spin, located at $x_{1}$, would decay exponentially with the separation $L$.
In order to show this property explicitly we need to estimate $\Delta$ appearing in Eq. (\ref{epsilon(omega)}).
To this end we first rewrite this equation as a regular Fourier transform and obtain
\begin{widetext}
\begin{equation}\label{e(t)}
\epsilon^{[h]}\left(t\right) = S\left(t\right)\cdot\frac{\Delta}{\delta\sqrt{2\pi}\sqrt{t_{0}^{2}\!-\!\left(2t\!+\!t_0\right)^{2}}}\!\left(\!e^{\frac{i}{\delta^{2}}\left[\frac{2t+t_0}{t_{0}}\cosh\left[A\right]+i\sqrt{1-\left(\frac{2t+t_0}{t_{0}}\right)^{2}}\sinh\left[A\right]\right]}\!+\!e^{\frac{i}{\delta^{2}}\left[\frac{2t+t_0}{t_{0}}\cosh\left[A\right]-i\sqrt{1-\left(\frac{2t+t_0}{t_{0}}\right)^{2}}\sinh\left[A\right]\right]}\right)\!,
\end{equation}
\end{widetext}
where
\begin{equation}
S\left(t\right)=\left\{ \begin{array}{ccc}
1 & , & -t_{0}\leq t \leq 0\\
0 & , & \text{else}
\end{array}\right..
\end{equation}
The singularity at $t=-t_0,0$ will disappear after the convolution with $h\left(t\right)$ which has been discussed in section III.
Therefore the function $\epsilon^{[h]}\left(t\right)$ will obtain its maximum at $t=-\frac{1}{2}t_0$ where it will be proportional to $\Delta \exp\left(\frac{1}{\delta^{2}}\sinh\left[A\right]\right)$.
In order for the perturbative expansion presented in Eq. (\ref{psi final}) to be justified, we require $\epsilon\left(t\right)\sim1$, hence
\begin{equation}
\Delta\sim e^{-\frac{1}{\delta^{2}}\sinh\left[A_\text{max}\right]}.
\end{equation}
Next, using the relation
\begin{eqnarray}
\delta^{2} & \ll & \frac{1}{\omega_{c}t_{0}\cosh\left[A_\text{max}\right]}\nonumber \\
 & \sim & \frac{1}{\omega_{c}T},
\end{eqnarray}
we get
\begin{eqnarray}
\Delta&\sim& e^{-\omega_{c} T\sinh\left[A_\text{max}\right]}\nonumber \\
&\sim& e^{-\frac{\omega_c L^2}{t_0}}.
\end{eqnarray}
The probability to postselect the spins as required for generating the remote field state is proportional to $\Delta^2$, therefore it decays generally like $P\sim \exp(-\frac{\omega_{c}L^2}{t_{0}})$.

The finiteness of the superoscillatory domain gives rise to an infidelity, $\eta \sim \int_{\omega_c}^{\infty}\frac{1}{\omega_c}\left|\tilde{F}\left(\bold{k}\right)\right|^2 d^d\bold{k}$.
Inverting the latter functional relation to $\omega_c=\omega_c(\eta)\equiv 1/g(\eta)$, we get the relation
\begin{equation} \label{P as a function of eta}
P\sim e^{-\frac{L^2}{g\left(\eta\right)t_{0}}},
\end{equation}
which describes the interplay between the success probability $P$ and the infidelity $\eta$.
When $\tilde{F}\left(\bold{k}\right)$ decays as a power law, $g\left(\eta\right)$ behaves like a power law as well,
and when $\tilde{F}\left(\bold{k}\right)$ decays exponentially $g(\eta) \sim 1/\text{ln}\left(1/\eta\right)$.
The decay of the success probability is therefore exponential with the separation $L$ - a feature that seems independent of the remotely generated function's shape. This feature could have been anticipated since the same exponential decay also characterizes the decay of vacuum entanglement between separated regions \cite{Reznik2005,Marcovitch2009}.

It is interesting to examine the sensitivity, or the tolerance, of the process to the effect of noise.
The key feature that leads to our results is related to the superoscillatory nature of the window function $\epsilon(t)$.
Let us consider the effect of adding noise to this function.
We could expect a correction of the form $\epsilon\left(t\right)\rightarrow\epsilon\left(t\right)+\nu\left(t\right)$, where $\nu\left(t\right)$ is some noise, and hence, the superoscillatory function receives an additive correction
$\tilde{\epsilon}\left(\omega\right)\rightarrow\tilde{\epsilon}\left(\omega\right)+\tilde{\nu}\left(\omega \right)$.
The effect of the noise may dominate the spin-field interaction unless $\nu$ is small enough.
An $\epsilon\left(t\right)\sim 1$ superoscillatory window function leads to an effect of amplitude as small as $\sqrt{P}\sim \exp(-\frac{\omega_{c}L^2}{t_{0}})$.
There is no reason to expect a similar suppression effect for the noise.
It therefore follows, that the present approach is only able to tolerate noise of amplitude $\nu< \nu_c\sim \exp(-\frac{\omega_{c}L^2}{t_{0}})$.
For $\nu>\nu_c$, the postselection of the spin(s) will generate a certain (random) field state.
In this case, since a typical $\tilde{\nu}(\omega)$ is not superoscillatory, the generated field state will generally live inside the future light-cone of $O_1$.

\section{VI. Relation to the Reeh-Schlieder theorem}
We now recall that our realization of remote preparation of field states was motivated from the Reeh-Schlieder theorem, which has been briefly discussed in the introduction.
According to this theorem, the set of field states generated from the vacuum by applying polynomials of the field operator in any fixed open region $O_1$ is dense in the whole Hilbert space $\mathcal{H}$.

Following a constructive approach, we have presented a method for realizing the sort of RSP described by the Reeh-Schlieder theorem: for every desired field state, $|\Psi\rangle$, we found a set of $N$ spins at $\bold{r}_i\in O_1$, certain local spins-field interactions for $-t_0<t<0$ and a spins' state $|D_f\rangle$, which at $t=0$ can be postselected.
Once the spins are postselected in this state we can assure that a field state, $|\Phi\rangle$, has been generated.
By taking the parameter $\delta$ to be arbitrarily small, one can set the window function, $\tilde{\epsilon}\left(\omega'\right)$, to be arbitrarily close to the desired window function, $\tilde{\epsilon}_\text{des}\left(\omega'\right)$, over an arbitrarily large domain $[0,\omega_c\left(\delta\right)]$.
Therefore, the generated field state can be made arbitrarily close to the desired field state, i.e., $|\langle \Psi|\Phi \rangle| >1-\eta\left(\delta\right)$.
While the success probability decays as the fidelity grows (since $\eta$ decreases as $\omega_c\left(\delta\right)$ grows) it always remains non-zero.

Since this method generates (with non-zero success probability) states which approximate any desired state arbitrarily well, the set of states which can be generated using this method is dense in the Hilbert space, hence it can be regarded as a constructive proof of the Reeh-Schlieder theorem.

While the Reeh-Schlieder theorem is restricted to standard QFT, this constructive approach may provide a glimpse to the process of RSP beyond the framework of standard QFT where new limitations are imposed due to the unknown physics at the Planck scale.
Adding a frequency cutoff to our model implies that while RSP is still possible in principle, the fidelity and the maximal separation between the operating region and the target region become restricted.

\section*{VII. Summary}
In this article we have provided and analysed a method for realizing remote preparation of relativistic quantum field states.
The mechanism that enables this task suggests that the phenomenon of superoscillations is fundamentally related to the Reeh-Schlieder theorem.
We believe that the suggested fundamental relation between the phenomenon of superoscillations and generalized quantum information tasks, such as remote state preparation, could open up new ways for studying the implications of quantum information theory within relativistic QFT.

\section*{Acknowledgements}
The authors would like to thank Erez Zohar for helpful discussions.
BR acknowledges the Israel Science Foundation.

\section*{Appendix: Generating an arbitrary one--particle field state in $1+1$ dimensions}

\begin{figure}
\includegraphics[scale=0.34]{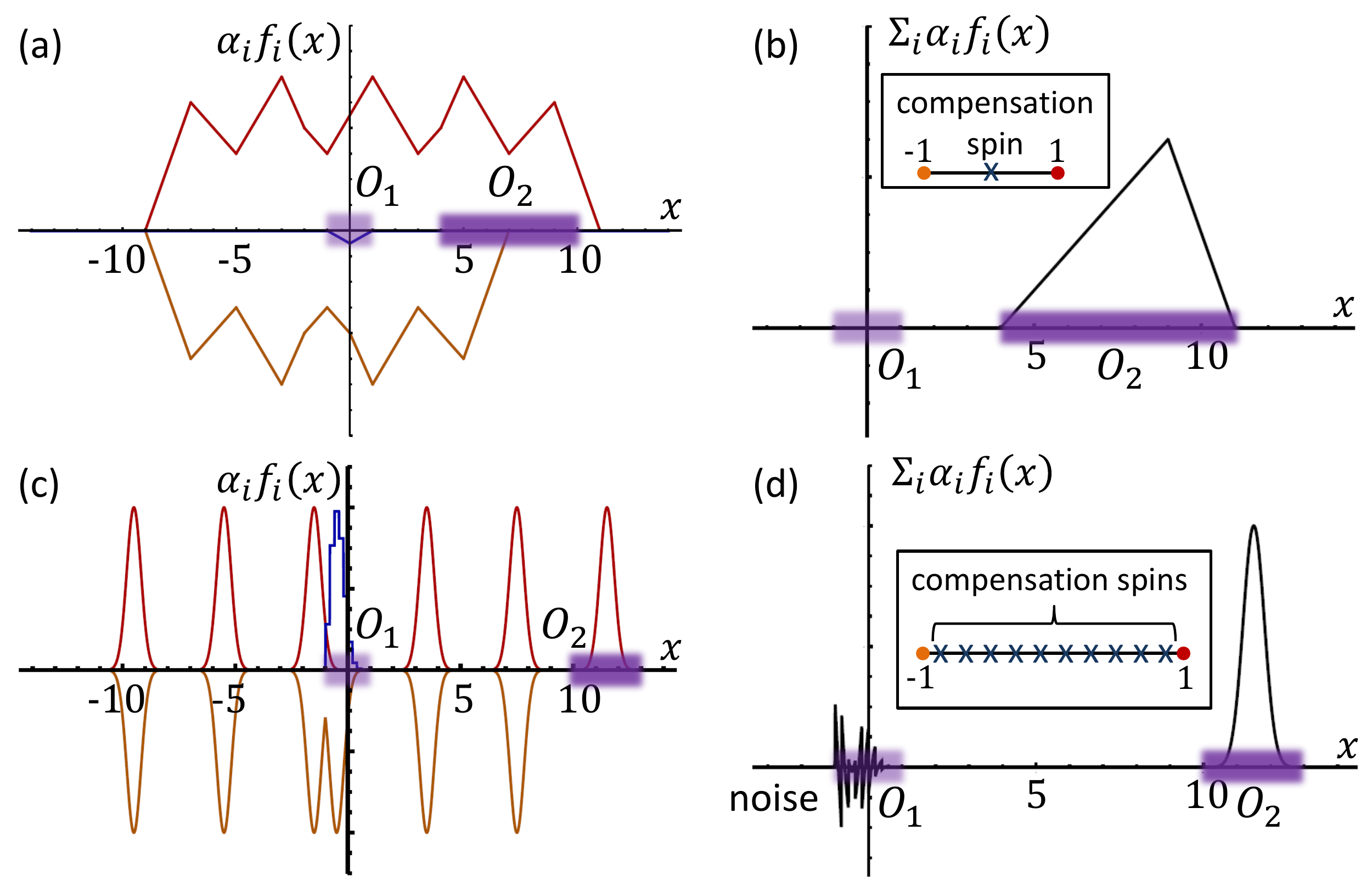}
\caption{(Color online) Schematically approximating selected field states in $1+1$ dimensions using arrays of spins.
The spins are set such that $d_{1}$ and $d_{2}$ are located at the end points of $O_1=\left[-1,1\right]$ and compensation spins are evenly spread in between.
(a) Exactly generating a desired field state using $3$ spins.
The functions $\alpha_{1}f_{1}\left(x\right)$ (orange), $\alpha_{2}f_{2}\left(x\right)$ (red) and $\alpha_{3}f_{3}\left(x\right)$ (blue) are presented.
Note that each of these functions is invariant under reflections around the position of the corresponding spin.
(b) A plot of $\protect\overset{3}{\protect\underset{i=1}{\sum}}\alpha_{i}f_{i}\left(x\right)$.
(c) Approximating another desired field state using $12$ spins.
The functions $\alpha_{1}f_{1}\left(x\right)$ (orange), $\alpha_{2}f_{2}\left(x\right)$ (red) and $\protect\overset{11}{\protect\underset{i=2}{\sum}} \alpha_{i}f_{i}\left(x\right)$ (blue) are presented.
(d) A plot of $\protect\overset{12}{\protect\underset{i=1}{\sum}}\alpha_{i}f_{i}\left(x\right)$.
In this case it is not possible to exactly generate the desired field state using a finite array of spins.
Note, however, that the error could be made arbitrarily small by increasing the number of compensation spins.
That would be at the expense of exponentially decreasing the
chances of success.}
\label{Detectors Distribution}
\end{figure}

Case study: In this appendix we illustrate the process of RSP in $1+1$ dimensions. A single spin could generate the most general one--particle spherically symmetric state. Therefore, following the postselection, an array of such spins, all located inside the region $O_1$, will generate the state
\begin{equation}
\left|\Phi\right\rangle =\underset{i}{\sum}\alpha_{i}\int d^d\bold{r} f_{i}\left(\left|\bold{r}-\bold{r}_{i}\right|\right)\phi\left(\bold{r}\right)\left|0\right\rangle,
\end{equation}
where $\bold{r}_i$ is the position of the $i$'th spin. Thus, in order to prepare an arbitrary one--particle field state $\left|\Psi\right\rangle =\int d^{d} \bold{r} F\left(\bold{r}\right)\phi\left(\bold{r}\right)\left|0\right\rangle$, we need to set the spin weight functions $\alpha_{i}f_{i}\left(\left|\bold{r}-\bold{r}_{i}\right|\right)$ such that
\begin{equation}
\underset{i}{\sum}\alpha_{i}f_{i}\left(\left|\bold{r}-\bold{r}_{i}\right|\right)=F\left(\bold{r}\right).\label{alphai fi = f(r)}
\end{equation}
In the $1+1$-dimensional case, since the spherical symmetry reduces to discrete $\mathbb{Z}_2$ reflection symmetry, it is particularly easy to find $\alpha_{i}f_{i}\left(\left|x-x_{i}\right|\right)$ that satisfy the above condition, which now takes the form
\begin{equation}
\underset{i}{\sum}\alpha_{i}f_{i}\left(\left|x-x_{i}\right|\right)=F\left(x\right).\label{alphai fi = f(x)}
\end{equation}
Let us choose to put two spins at the points $x_1=a,x_2=-a$.
It is then easy to verify that the functions
\begin{eqnarray}
\alpha_1\!f_{1}\!\left(\xi\right)&\!=\!&\!\sum_{n}\!\left(F\!\left(\!\xi\!+\!\left(4n\!+\!1\right)\!a\!\right)\!-\!F\left(\!-\!\xi\!-\!\left(4n\!+\!3\right)\!a\!\right)\!\right)\\
\alpha_2\!f_{2}\!\left(\xi\right)&\!=\!&\!\sum_{n}\!\left(F\!\left(\!-\!\xi\!-\!\left(4n\!+\!1\right)\!a\!\right)\!-\!F\left(\xi\!+\!\left(4n\!+\!3\right)\!a\!\right)\!\right)
\end{eqnarray}
solve Eq. (\ref{alphai fi = f(x)}) everywhere except in the segment $[-a,a]$.
Here we have implicitly assumed that the given $F(x)$ is fast decreasing at $|x|\rightarrow\infty$.
In order to correct the field in the domain $[-a,a]$ we add $N-2$ extra ``compensation'' spins inside this region.
Each of these $N-2$ spins would eliminate the field state in its neighbourhood, and in the limit $N\gg1$ they will converge to completely cancel out the field state in $\left[-a,a\right]$.
Thus we are left with the desired field state $\left|\Psi\right\rangle =\int dx F\left(x\right)\phi\left(x\right)\left|0\right\rangle $.
In Fig. (\ref{Detectors Distribution}) we demonstrate this method.

\bibliography{ref}

\end{document}